\documentstyle[aps,multicol,epsf,epsfig]{revtex}
  
\begin{document}

\newcommand{\np}{\newpage}
\newcommand{\xsize}{\epsfxsize=14.0cm}

\draft

\title{Percolation Threshold, Fisher Exponent,\\
       and Shortest Path Exponent for 4 and 5 Dimensions}

\author{Gerald Paul,$^1$\thanks{electronic address:
gerryp@bu.edu} Robert M. Ziff,$^2$ and H. Eugene Stanley$^1$}

\address{$^1$Center for Polymer Studies and Department of Physics\\
Boston University, Boston, MA 02215\\
$^2$Center for Theoretical Physics and Department of Chemical Engineering\\
University of Michigan, Ann Arbor, MI 48109-2136}

\date{pzs.tex ~~~ 10 January 2001~~~ }

\maketitle

\begin{abstract}

We develop a method of constructing percolation clusters that allows us
to build very large clusters using very little computer memory by
limiting the maximum number of sites for which we maintain state
information to a number of the order of the number of sites in the
largest chemical shell of the cluster being created. The memory required
to grow a cluster of mass $s$ is of the order of $s^\theta$ bytes where
$\theta$ ranges from 0.4 for 2-dimensional lattices to 0.5 for 6- (or
higher)-dimensional lattices.  We use this method to estimate 
$d_{\mbox{\scriptsize min}}$, the exponent relating the minimum path
$\ell$ to the Euclidean distance $r$, for 4D and 5D hypercubic
lattices. Analyzing both site and bond percolation, we find
$d_{\mbox{\scriptsize min}}=1.607\pm 0.005$ (4D) and
$d_{\mbox{\scriptsize min}}=1.812\pm 0.006$ (5D). In order to determine
$d_{\mbox{\scriptsize min}}$ to high precision, and without bias, it was
necessary to first find precise values for the percolation threshold,
$p_c$: $p_c=0.196889\pm 0.000003$ (4D) and $p_c=0.14081\pm 0.00001$ (5D)
for site and $p_c=0.160130\pm 0.000003$ (4D) and $p_c=0.118174\pm
0.000004$ (5D) for bond percolation.  We also calculate the Fisher
exponent, $\tau$, determined in the course of calculating the values of
$p_c$: $\tau=2.313\pm 0.003$ (4D) and $\tau=2.412\pm 0.004$ (5D).

\end{abstract}

\begin{multicols}{2}

\section {Introduction}

Percolation is a standard model for disordered systems
\cite{Stauffer,BundeHavlin}. In percolation systems, sites or bonds on a
lattice are populated with probability $p$. The value of $p$ at which
infinite clusters are formed is known as the critical probability or
percolation threshold $p_c$. The shortest path exponent,
$d_{\mbox{\scriptsize min}}$, is defined by the relation
\cite{BundeHavlin,Grassberger92a,Grassberger92b}
\begin{equation}
\label{e1}
\langle\ell\rangle\sim r^{d_{\mbox{\scriptsize min}}},
\end{equation}
where $r$ is the Euclidean distance between two sites on a cluster and
$\ell$ is the length of the shortest path traveling along occupied sites
and bonds in the percolation cluster (``chemical distance''). We can
also write
\begin{equation}
\label{e2}
\langle r\rangle\sim\ell^z,
\end{equation}
which defines the exponent $z=1/d_{\mbox{\scriptsize min}}$. With the
exceptions of $d\geq 6$ (where $z=1/2$) and $d=1$ (where $z=1$), $z$ is
not known exactly.  The most common method of determining $z$
numerically (and the one we will use) is to grow clusters, calculating
the average distance $\langle r\rangle$ of sites in the cluster from the
seed of the cluster as a function of chemical distance, $\ell$, from the
seed. In order that finite size effects do not play a role, the lattice
must be large enough such that the clusters which are grown do not reach
the boundaries of the lattice.

Because corrections-to-scaling decrease with increasing $\ell$, the
larger the value of $\ell_{\mbox{\scriptsize max}}$ (the value of $\ell$
at which we stop the growth), the more accurately we can estimate $z$.
The limitations on the size, $\ell_{\mbox{\scriptsize max}}$, to which
the clusters can be grown have been the computer memory available for
the simulation and the computer processing power needed to build these
clusters. The method of ``data blocking'' \cite{Ziff1984,Lorenz} has
helped ameliorate the need for large amounts of memory. In this method,
the lattice is logically divided into blocks; memory for a block is not
allocated until the lattice grows into that block.  The data blocking
method has been used recently to obtain precise estimates for the
percolation threshold and associated exponents for bond and site
percolation on a number of lattices
\cite{Lorenz,StaufferZiff}. Ultimately, however, although sufficient
computer power is available to build larger clusters, the cluster size
is limited by the amount of memory available. This becomes particularly
true as the dimension of the lattice $d$ increases since at criticality
the cluster becomes less dense as $d$ increases \cite{text1}. To reach
the same cluster mass or $\ell_{\mbox{\scriptsize max}}$, we must have
larger lattices.

In this paper we describe a method of constructing clusters which
dramatically reduces the memory requirements needed to grow large
clusters relative to previous methods. Using this method of building
large clusters, we estimate $z$ for hypercubic lattices in 4, and 5,
dimensions.  The study of critical properties in higher dimensions is
important because one can use the results to test relations which are
conjectured to hold in all dimensions (hyperscaling relations) and
exponents which are believed to be the same in all dimensions
(superuniversal exponents). The current best estimates of
$d_{\mbox{\scriptsize min}}$ for 4 and 5 dimensions, 1.5 and 1.8,
respectively \cite{Stauffer}, are of relatively low precision compared
to the estimates available in 2 and 3 dimensions $1.1307\pm 0.0004$ and
$1.374\pm 0.006$, respectively \cite{BundeHavlin,Grassberger92a}.

\section{Cluster Generation}

One method of cluster generation is the Leath method \cite{Leath}.  In
this method a site is chosen as the seed site of the cluster.  Using a
random number generator and a given bond occupation probability, one
determines whether the bonds connected to the seed site are occupied or
not \cite{bondpercolatonused}. If a bond is occupied, the site to which
this bond connects is considered to be part of the cluster and becomes a
``growth site.''  These sites are at chemical distance of unity from the
seed site; all sites at the same chemical distance, $\ell$, from the
seed site are considered to be in ``chemical shell $\ell$.'' The process
is then repeated for each of these growth sites with the next set of
growth sites being at chemical distance 2 from the seed site. The
cluster continues to grow until the growth stops naturally, the growth
is terminated by the sides of the $d$-dimensional lattice of edge $L$,
or the maximum chemical distance, $\ell_{\mbox{\scriptsize max}}$, is
reached.

We use the Leath method to construct clusters, but we keep track of
which bonds are occupied and which sites have been visited by a method
different from that traditionally used. Traditionally, this state
information is stored in an array of size equal to the number of lattice
sites. In the data blocking method, memory usage can be improved by
allocating blocks in this array dynamically.  Vollmayr \cite{Vollmayer}
eliminated the use of this array, storing status of visited sites in a
data structure thus reducing memory requirements to grow a cluster of
mass $s$ to $O(s)$. We extend the approach of Ref.~\cite{Vollmayer}
further, reducing the memory required to $O(s^\theta)$ where $\theta$
ranges from 0.4 for 2-dimensional lattices to 0.5 for 6- (or
higher)-dimensional lattices.

To see how this can be done, we first consider the uses of this state
information:

\begin{itemize}

\item[{(a)}] {\it Occupancy status\/}: Information concerning whether a
site/bond is occupied is maintained so that it is the same, independent
of when in the growth process it is accessed. For example, we would not
generate a cluster with the proper statistics if we treated a bond as
occupied during one stage of the cluster growth and then treated it as
empty during a later stage.

\item[{(b)}] {\it Visited status\/}: Information concerning whether a
site has been visited or not is maintained in order that (a) we do not
multiply count the presence of a site in the cluster and (b) we do not
retrace our steps during cluster generation, causing the growth process
to never end.

\end{itemize}

\subsection{Occupancy Status}

We address the need to maintain information about whether a bond is
occupied or not by using a random number generation scheme in which the
random number associated with a bond is determined by the location of
the bond in the lattice and the orientation of the bond. This is done by
first assigning a unique number $n$ to any {\it site\/} in the lattice as follows:
Let ($x_1,x_2,x_3\ldots x_d$) be the coordinates of the site in the
lattice, and let ($L_1,L_2,L_3\ldots L_d$) be the lengths of the sides
of the lattice. Then
\[
n(x_1,x_2,x_3\ldots x_d) =
\]
\begin{equation}
\label{e3}
[(\{[(x_1 L_2)+x_2] L_3\} + x_3\ldots)L_d+x_d]
\end{equation}
assigns a unique number to any site in the lattice. We assign a unique
number, $n'$, to any {\it bond\/} in the lattice by defining
\begin{equation}
\label{e4}
n'(x_1,x_2,x_3\ldots x_d,o) = [n(x_1,x_2,x_3\ldots x_d) d]+o,
\end{equation}
where $o$ is the orientation of a bond attached to site $x$ (assuming
values 0 to $d-1$).

Furthermore we want to assign unique numbers to bonds over many
different realizations. We then define
\[
n''(x_1,x_2,x_3\ldots x_d,o,m) =
\]
\begin{equation}
\label{e5}
[n'(x_1,x_2,x_3\ldots x_d,o) M]+m,
\end{equation}
where $m$ is the number of the realization and $M$ is the maximum number
of realizations we plan to create.

We then generate a 64-bit random number, $R$, using an encryption-like
algorithm $f(n'')$ \cite{NumericalRecipes} using $n''$ as its input,
\begin{equation}
\label{e6}
R=f(n'').
\end{equation}
A bond is occupied if $R>2^{64}p$. In practice, because for large
lattices and a large number of realizations $n''$ is greater than
$2^{64}$, the maximum size of the input to the random number algorithm,
we actually determine the random number in two steps,
\begin{equation}
\label{e7}
\bar R=f(\{[f(n)d]+o\}M+m).
\end{equation}
That is, we first create an intermediate random number based only on the
coordinates of the bond and then create the final random number based on
the intermediate random number, the orientation of the bond and the
realization number. Using the test described in \cite{Ziff98}, we
confirm that, within statistical error, our algorithm generates unbiased
random numbers.  This test is important because there is only a small
difference between the inputs to the random number generator for
neighboring sites. Any correlations between the outputs would cause
incorrect results \cite{Ziff98}. The generation of random numbers using
Eq.~(\ref{e7}) is slower than congruence or shift register techniques
\cite{Ziff98} but is somewhat compensated by eliminating the processing
done to store and access bond state when maintained in an array. In any
case, the net effect of using this approach is about a factor of 5
increase in calculation time because of the slowness of the
encryption-like random number generator that we used.

\subsection{Visited Status}

We address the need to maintain information about whether a site has
been visited or not by storing information about visited sites in a data
structure. Each entry in the data structure contains the coordinates of
the site, the chemical shell of the site, and a bit map with one bit for
each direction from which the site can be visited. The data structure
can be accessed as a ``circular list'' (last-in-first-out queue) so
entries can be added and deleted. Since a site can be visited from
different directions, we must ensure that a site is counted only once
and that backtracking does not occur. To accomplish this, before adding
a site to the list of growth sites, we first check to see if it is
already in the list. 

\begin{itemize}

\item
If it is already on the list, we do not add a new entry but, in the
entry for the site already in the list, we do set the bit corresponding
to the orientation of the connected bond which was traversed to visit
the site. 

\item
If it is not in the list, we add it (storing the coordinates and
chemical length and setting the bit corresponding to the direction from
which the site was visited). 

\end{itemize}

When we are about to process the entry for a growth site, we only count
the site once in the mass of the cluster, and only attempt to grow the
cluster in directions other than those from which the site was
visited. In this way we avoid backtracking along already traveled paths.
If the data structure had to be searched sequentially every time we were
about to create a growth site, the time needed would make this approach
impractical. In \cite{Vollmayer}, the data structure was maintained as a
binary tree in order to reduce search time. We use the faster ``hash
table'' method \cite{Knuth} to access entries for the visited sites.
 
The hashing technique works as follows: A key, $K$, is associated with
each entry of the data structure. We use a function
$h(K)$ to map the key into a ``slot'' at offset $h(K)$ in a ``hash
table''. If the slot in the table is not already used, we store the
number or address of the entry in this slot; if the slot is used (this is
referred to as a ``collision'') we add the entry to a chain
of entries all of which map to the same value $h(K)$. Ideally the function
$h(K)$ maps the keys uniformly over the slots in the
table so we obtain few long chains. If we use a hash table of 
size $M=2^m$, where $m$ is an integer and choose $K$ as
the unique number, $n$, of the site, an effective hashing function is
\cite{Knuth}
\begin{equation}
\label{e1x}
h(n)={1\over 2^{w-m}}\left[(n~C)~\mbox{mod}~2^{w}\right],
\end{equation}
where w is the word size(in bits) of our computer and the hash constant,
$C$ is the least significant $w$ bits of the product of $2^w$ and the
``golden ratio'', $(\sqrt 5-1)/2$.  Thus $h(n)$ yields the upper $m$ bits (shift right $w-m$ bits) of the result of
taking the lower $w$ bits of the product of the unique site number and
the hash constant, $C$. We implement the ability to chain
entries in the data structure by defining another field in the data
structure entry which serves as a chain pointer field. To find an entry
in the data structure for a site, we calculate the unique site number
using Eq.~(\ref{e3}), find the offset in the hash table using
Eq.~(\ref{e1x}), and then walk the chain of entries to find the entry
with the desired coordinates.  If we make the size of the hash table
equal to the size of the
site data structure, we find the average number of hash ``collisions''
to be less than 2 so we can determine if a site has been visited very
efficiently.

\begin{figure}
%\centerline{
\epsfxsize=14.0cm
\epsfbox{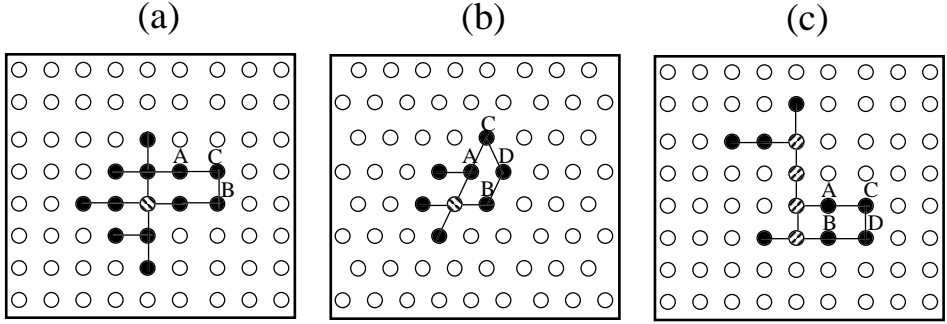}
%}

%\psfig{file=f0.ps,width=14cm,clip=,angle=-90}

\caption{Examples of cluster growth at the beginning of the population
of sites at chemical distance 3 from the seed site. The seed sites are
denoted by striped circles. (a) Example of a square lattice in which a
site, C, is multiply-visited from sites A and B. (b) Example of a
triangular lattice in which site C can be multiply-visited from sites A
and D and in which site D can be multiply-visited from sites B and
C. (c) Example in which the cluster is grown from multiple seeds. Site C
can be multiply-visited from sites A and D; site D can be
multiply-visited from sites B and C.}
\label{f1}
\end{figure}

This approach of keeping the status of visited sites in a special data
structure (not in the lattice array) applies to any lattice model. In
the case of growing percolation clusters we can further reduce the
amount of memory needed significantly. This is accomplished by
recognizing that a site which is multiply visited is done so during the
growth of a single chemical shell. This is the key insight that allows
us to reduce the memory requirement and can be confirmed by considering
the bonds adjacent to a site in a lower chemical shell: (i) an occupied
bond adjacent to a site in a lower chemical shell cannot be a path to
re-visit that site because we do not back-track and (ii) an unoccupied
bond adjacent to a site in a lower chemical shell cannot be on the path
to revisit that site. Sites in the same chemical shell can, however, be
visited by multiple paths as shown in Fig.~\ref{f1}(a). Thus we need
only keep state information about growth sites which themselves have not
yet been used to create entries for the next chemical shell. The number
of such sites at any point in the growth process will be of the order of
the size of the current chemical shell.

The discussion so far has been for hypercubic lattices. For these
lattices, we ensure that we did not double-count site or backtrack by
maintaining information about growth sites which themselves have not yet
been used to create entries for the next chemical shell and then
checking for duplicates. More generally (e.g., for triangular lattices),
the situation is a little more complicated as shown in the example in
Fig.~\ref{f1}(b). A similar situation is shown in Fig.~\ref{f1}(c),
where we grow a cluster from multiple seeds. To treat both types of
situation, we must maintain (i) state information about growth sites
which themselves have not yet been used to create entries for the next
chemical shell and (ii) state information about all sites in the
chemical shell previous to the one being built. Before we add a site to
the list of growth sites, we check if it is already present in the
previous shell; if it is, we do not add it.

The size of a chemical shell can be estimated as follows. The chemical
distance, $\ell$, scales with the mean Euclidean radius of the cluster,
$r$, as
\begin{mathletters}
\begin{equation}
\label{e8}
\ell\sim r^{d_{\mbox{\scriptsize min}}},
\end{equation}
while the cluster mass (the number of sites in the cluster), $s$, scales
as
\begin{equation}
\label{e8a}
s\sim r^{d_f},
\end{equation}
\end{mathletters}
where $d_{\mbox{\scriptsize min}}$ has values 1.13 and 2 for $d=2$ and
6, respectively \cite{BundeHavlin,Grassberger92a,Pike81,Herrmann88};
$d_f$, the fractal dimension of the cluster mass, has the exact values
$91/48=1.89$ and 4 for $d=2$ and 6, respectively
\cite{Stauffer,BundeHavlin}. Then
\begin{equation}
\label{e9}
s\sim\ell^{d_f/d_{\mbox{\scriptsize min}}},
\end{equation}
and
\[
ds\sim\ell^{(d_f/d_{\mbox{\scriptsize
min}})-1}d\ell=(s^{d_{\mbox{\scriptsize
min}}/d_f})^{(d_f/d_{\mbox{\scriptsize
min}})-1}d\ell=
\]
\begin{equation}
\label{e10}
s^{1-(d_{\mbox{\scriptsize
min}}/d_f)}d\ell.
\end{equation}
Setting $d\ell=1$, we find the size of the outermost chemical shell of a
cluster of mass $s$ scales as
\begin{equation}
\label{e11}
s_{\mbox{\scriptsize shell}}(s)\sim s^\theta
\end{equation}
where $\theta=1-d_{\mbox{\scriptsize min}}/d_f$.

The values of $\theta$ range from $\approx 0.4$ to 0.5 for $d=2$ to
$d=6$. Thus the size of the data structure to contain the visited status is
only of the order of the square root of the size of the cluster size
because we only store status for the largest chemical shell.

\section {Percolation Threshold and Fisher Exponent}

In order to determine $d_{\mbox{\scriptsize min}}$ to high precision and without bias, it
is necessary to first find values of the percolation threshold
substantially more precise than previously known (in most cases).  To
determine these thresholds we used the method of measuring cluster-size
statistics of individual clusters grown on large virtual lattices as
described in \cite{Lorenz}.  The data-blocking method \cite{Ziff1984}
used involves assigning memory to parts of the lattice only when the
cluster grows into it.  With the data-blocking method, like the hashing
method, a table is used to access a data structure but in this case, the
data structure entries represent blocks of sites instead of individual
sites; there are no collisions, but some memory is wasted. The advantage
of using the data-blocking method as opposed to the one proposed in this
work is that the state of all sites are recorded (as described in
appendix A), so it allows using a fast random number generator.  The
data blocking method method allows lattices of sufficient size to keep
finite-size effects under control, with sufficient speed to achieve good
statistics.  (The hashing method described in this paper could also have
been used for this calculation.)

In 4D, we use a virtual lattice of $(512)^4$ sites, broken up into
blocks of $(16)^4$ sites each.  In 5D, the virtual lattice of size
$(128)^5$ is divided into blocks of size $8^5$. The cluster-size cutoff,
$s_{\mbox{\scriptsize max}}$, is $2^{17} = 131,072$ and $2^{14}= 16,384$
for 4D and 5D, respectively. The threshold is determined as the value of
$p$ that leads to the cluster size distribution $n_s$ best following a
power-law $n_s \sim s^{-\tau}$.  Simulating about $10^8$ clusters for
each case, and using the data analysis techniques employed in
\cite{Ziff1984}, we find
\begin{equation}
\label{pcsitebond}
p_c=\cases{
0.196889\pm 0.000003 & [4D site]\cr
0.160130\pm 0.000003 & [4D bond]\cr
0.14081\pm 0.00001    & [5D site]\cr
0.118174\pm 0.000004 & [5D bond]
}.
\end{equation}
Also, for $\tau$ we find the values
\begin{equation}
\label{tausitebond}
\tau=\cases{
2.313\pm 0.003  & [4D] \cr
2.412\pm 0.0004 & [5D]
}.
\end{equation}
These results are more precise than some of the published values for
$p_c=0.16005\pm 0.00015$ \cite{Adler}, $0.1407\pm 0.0003$
\cite{Marck98}, and $0.11819\pm 0.00004$ \cite{Adler} for 4D bond, 5D
site, and 5D bond percolation, respectively and for $\tau=2.41$ for 5D
percolation; for 4D site percolation, Ballesteros et al.\cite{Ball}
found the comparably precise value $p_c = 0.196901\pm 0.000005$ (just
slightly higher than ours) and $\tau =2.3127\pm0.0007$.  All simulation
parameters and our results are summarized in Table I.

The precision of our results is sufficiently high that we expect that
statistical errors in $p_c$ will not have an effect on our value of
$d_{\mbox{\scriptsize min}}$.

\section{Shortest Path Exponent}

To calculate the shortest path exponent, we ran simulations at the
percolation thresholds found above.  We stopped cluster growth at
$\ell_{\mbox{\scriptsize max}}=2048$ for 4D bond and site percolation
and $\ell_{\mbox{\scriptsize max}}=1024$ for 5D bond and site
percolation. We simulated $73\times 10^6$, $39\times 10^6$, $105\times
10^6$, and $20\times 10^6$ realizations for 4D bond, 4D site, 5D bond,
and 5D site percolation, respectively.  During our simulations, we kept
track of the maximum and minimum lattice points to which our clusters
extended.  Using this information, we determined the size of the lattice
that we would have needed to build if we had been using conventional
memory techniques.  For $d=5$ the lattice would have had sides of length
$L=245$ resulting in approximately $900\times 10^9$ lattice sites
($\approx 1$TB memory); the actual memory used was less than $\approx
10^6$ (1MB), six orders of magnitude smaller.

\begin{figure}
\centerline{
\xsize
\epsfclipon
\epsfbox{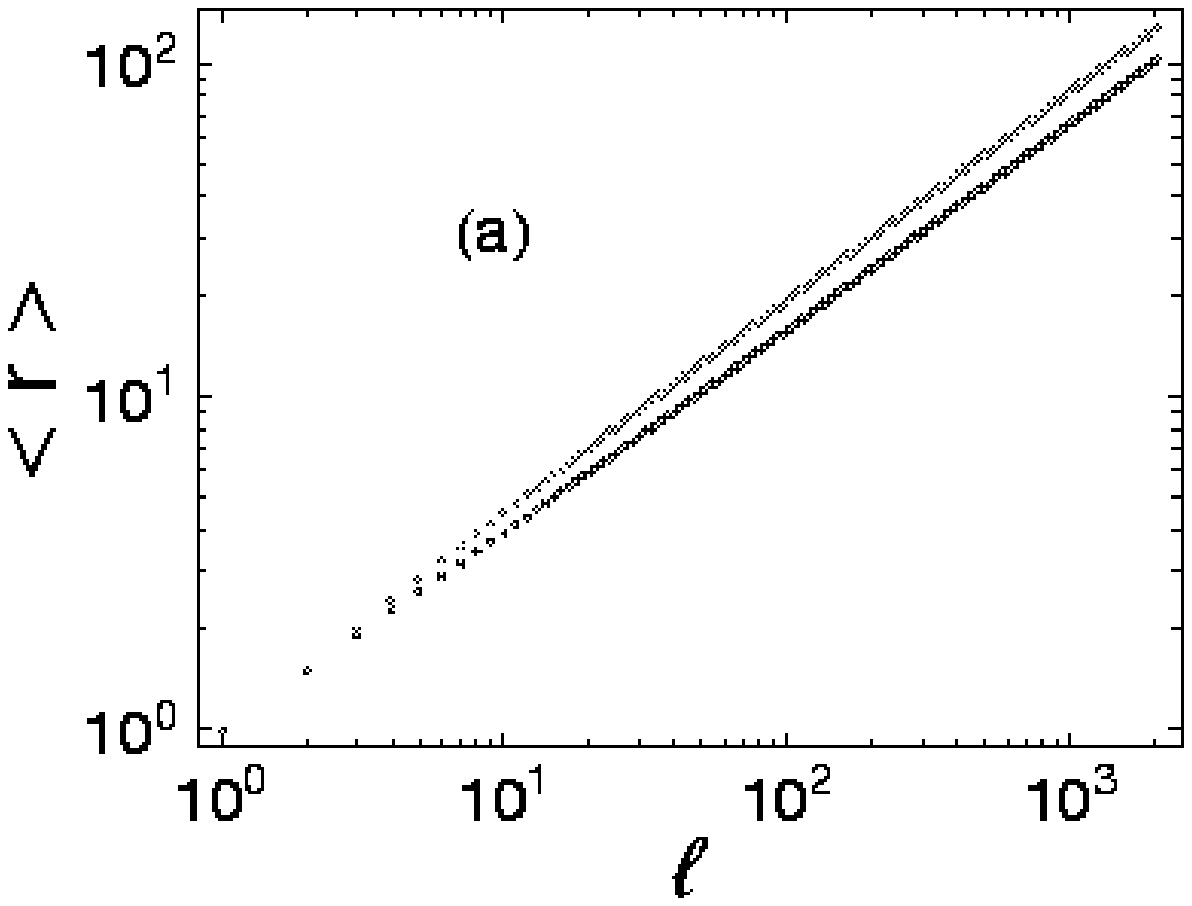}
}
\centerline{
\xsize
\epsfclipon
\epsfbox{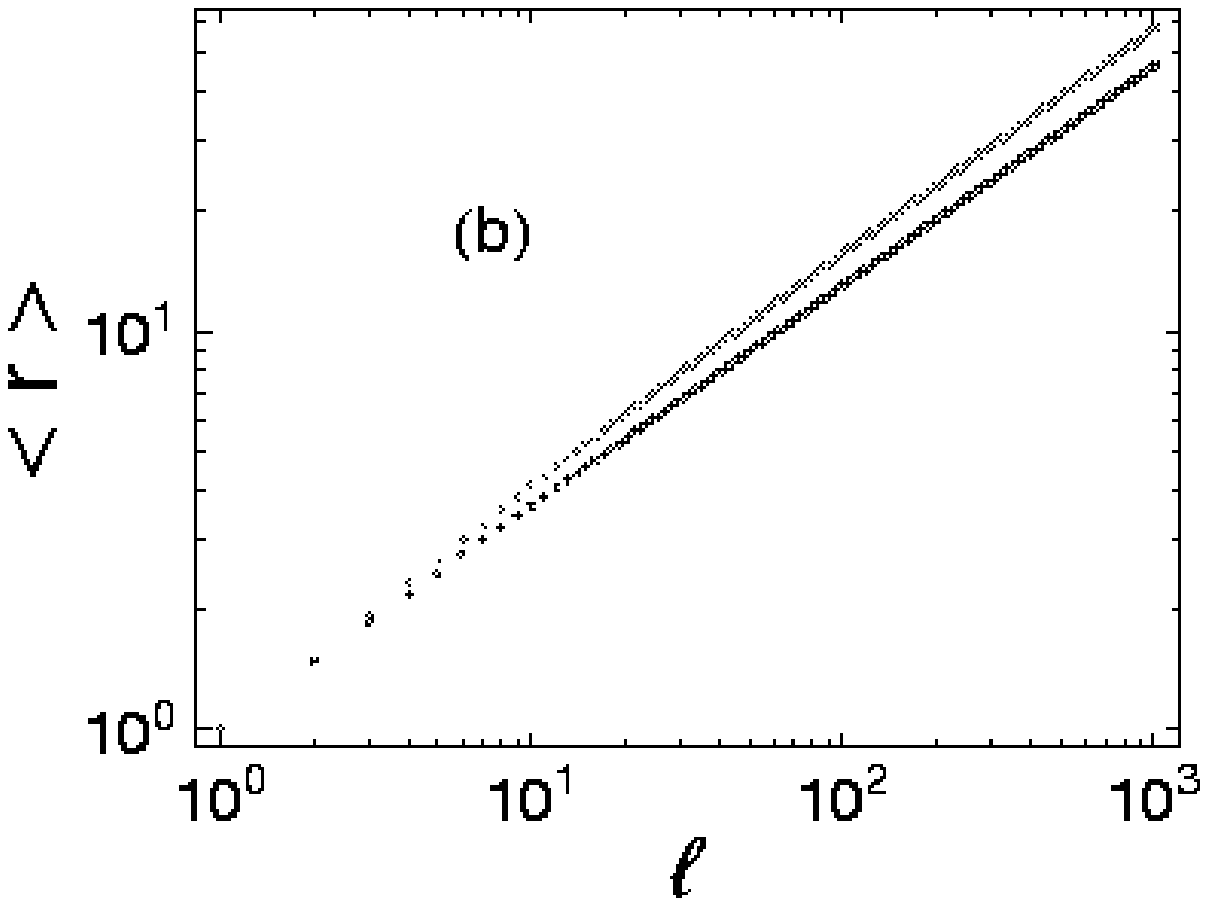}
}
\caption{Euclidean distance $\langle r \rangle$ versus chemical
distance $\ell$ for site percolation (upper line) and bond percolation
(lower line) for (a) 4D and (b) 5D. The slightly different apparent
slopes of the plots for bond and site cases are due to different values
of the correction-to-scaling parameters.}
\label{f2}
\end{figure}

Figure \ref{f2}a shows plots of $\langle r \rangle$ for 4D site and bond
percolation, while Fig.~\ref{f2}b shows plots of $\langle r \rangle$ for
5D site and bond percolation. While the plots resemble straight lines,
the effects of corrections-to-scaling are, in fact, considerable. One
customarily assumes that corrections-to-scaling have the functional form
\cite{Stauffer,BundeHavlin,Grassberger92a,Grassberger92b}
\begin{equation}
\label{e12}
\langle r \rangle\sim\ell^z(1+A\ell^{-\Delta}+\cdots),
\end{equation}
where the constant $A$ depends on the dimension, lattice type and
percolation type (bond or site) but the exponent $\Delta$ depends only
on dimension. Let
\begin{equation}
\label{e13}
h(\ell)\equiv{\langle r \rangle\over\ell^{z'}}\sim\ell^{z-z'}(1 +
A\ell^{-\Delta}+\cdots), 
\end{equation}
where $z'$ is an estimated value of $z$. If $\ell_{\mbox{\scriptsize
max}}$ were infinitely large, we could determine $z$ as the value of
$z'$, which results in a plot of $h(\ell)$ which asymptotically
approaches a constant (i.e., has zero slope as $\ell\to\infty$);
however, since $\ell_{\mbox{\scriptsize max}}$ is finite, we may obtain
misleading results if we determine $z$ in this manner. Nevertheless, we
can use this approach to determine bounds on $z$.

\begin{figure}
\centerline{
\xsize
\epsfclipon
\epsfbox{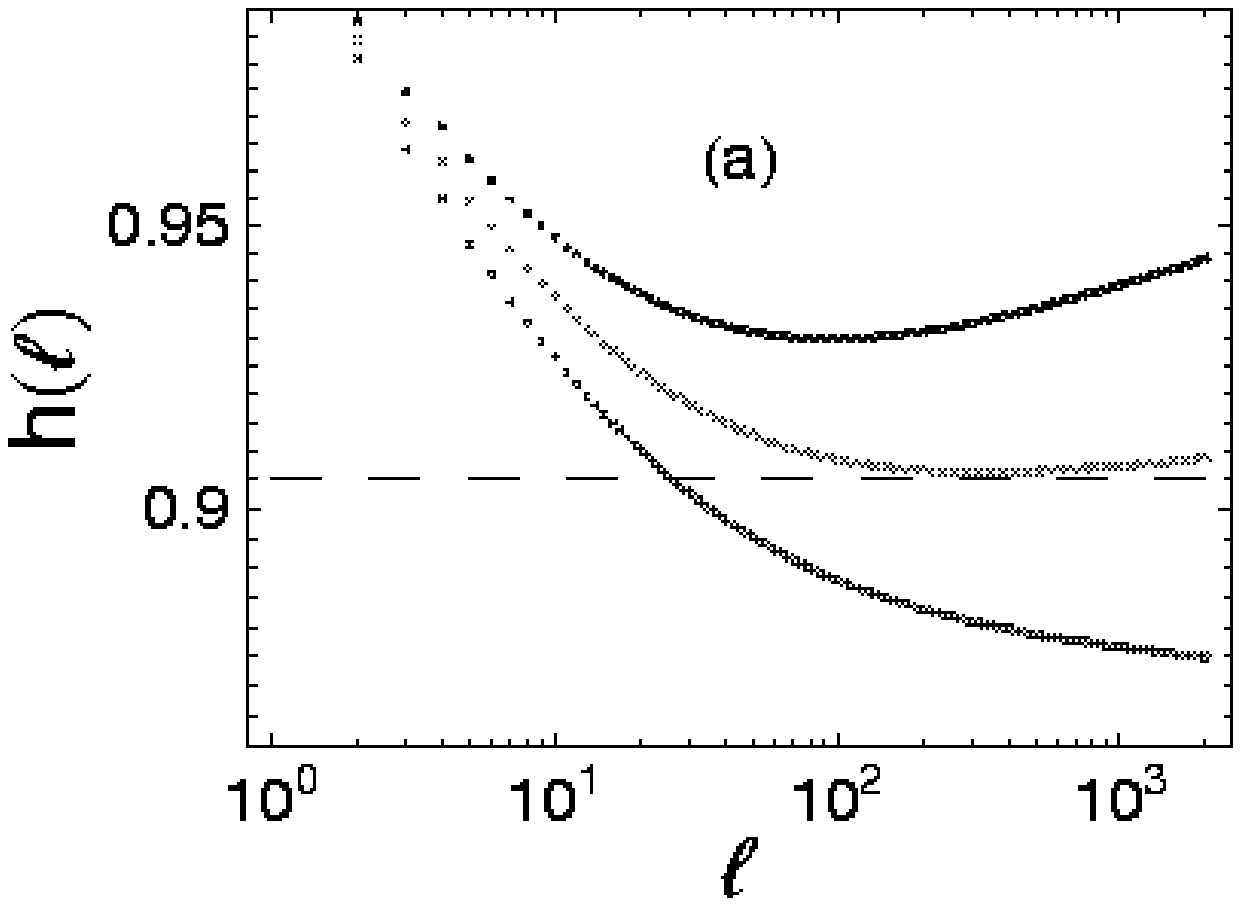}
} 

\centerline{
\xsize
\epsfclipon
\epsfbox{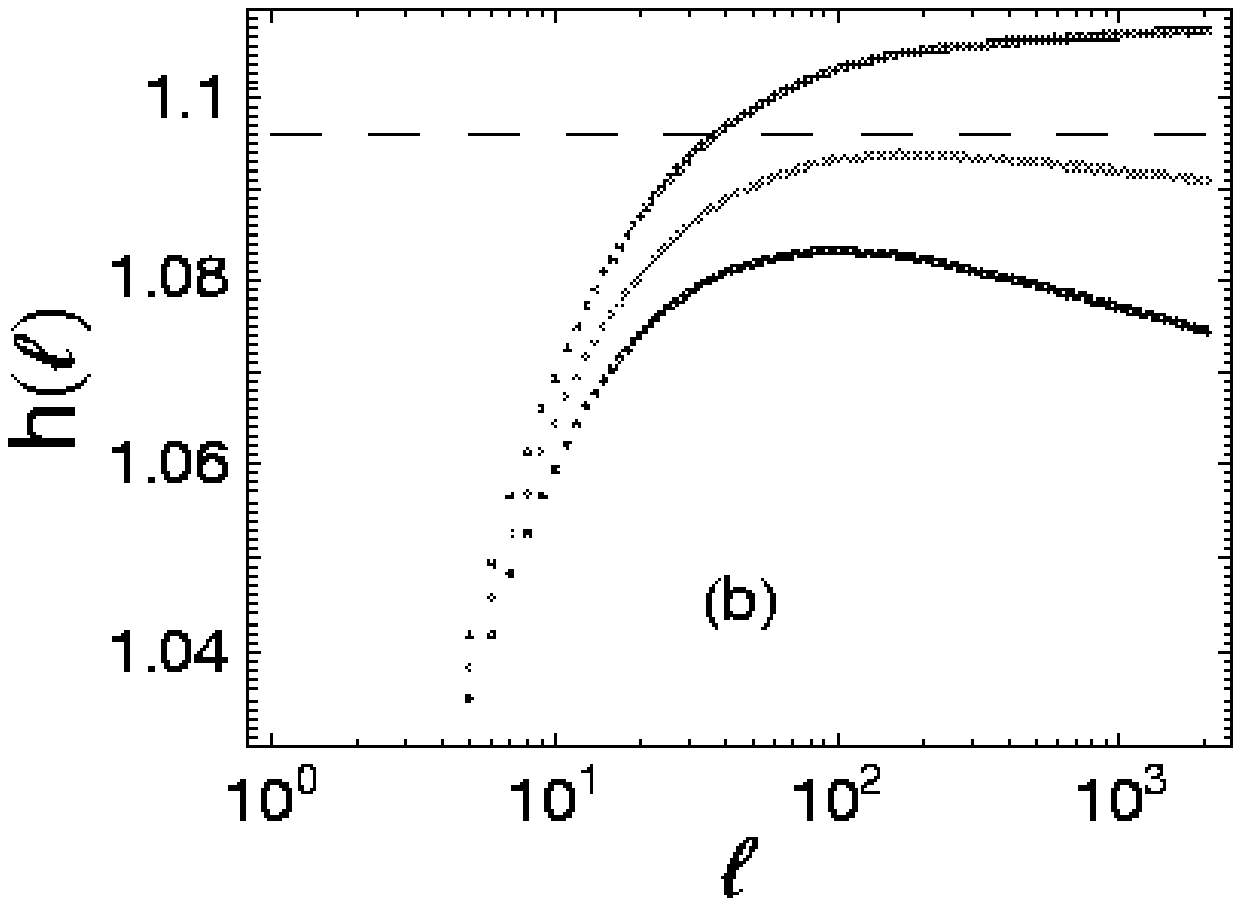}
} 
\caption{$h(\ell)\equiv\langle r \rangle/\ell^z$ versus $\ell$ for (a) 4D
bond percolation for values of (from top to bottom) $z'=0.615$, 0.620
and 0.625 (b) 4D site percolation for values of (from top to bottom)
$z'=0.623$, 0.625, and 0.627. The dashed horizontal lines are provided
as guides to the eye to allow one to better see that, for large $\ell$,
the middle plots of $h(\ell)$ in (a) and (b) are increasing and
decreasing, respectively.}
\label{f4}
\end{figure}

To see how this is accomplished, first consider Fig.~\ref{f4}a, in which
we plot $h(\ell)$ for 4D bond percolation for various values of
$z'$. From this figure and Eq.~(\ref{e13}) it is clear that $A$ is
positive. Hence, we know that if for large $\ell$ the slope of $h(\ell)$
becomes an increasing function, the leading power-law term $\ell^{z-z'}$
will dominate because $z>z'$. Thus a lower bound on $z$ is that value of
$z'$ at which $h(\ell)$ asymptotically becomes an increasing
function. From Fig.~\ref{f4}a this value is 0.620.

We can proceed similarly by considering site percolation in 4D, plotting
$h(\ell)$ for 4D site percolation for various values of $z'$ in
Fig.~\ref{f4}b. From these plots it is clear that $A$ for bond
percolation is negative. Hence we know that if for large $\ell$ the
slope of $h(\ell)$ becomes a decreasing function, we are seeing the
leading power-law term $\ell^{z-z'}$ dominate because $z<z'$. Thus an
upper bound on $z$ is that value of $z'$ at which $h(\ell)$
asymptotically becomes a decreasing function. From Fig.~\ref{f4}b this
value is 0.625.

\begin{figure}
\centerline{
\xsize
\epsfclipon
\epsfbox{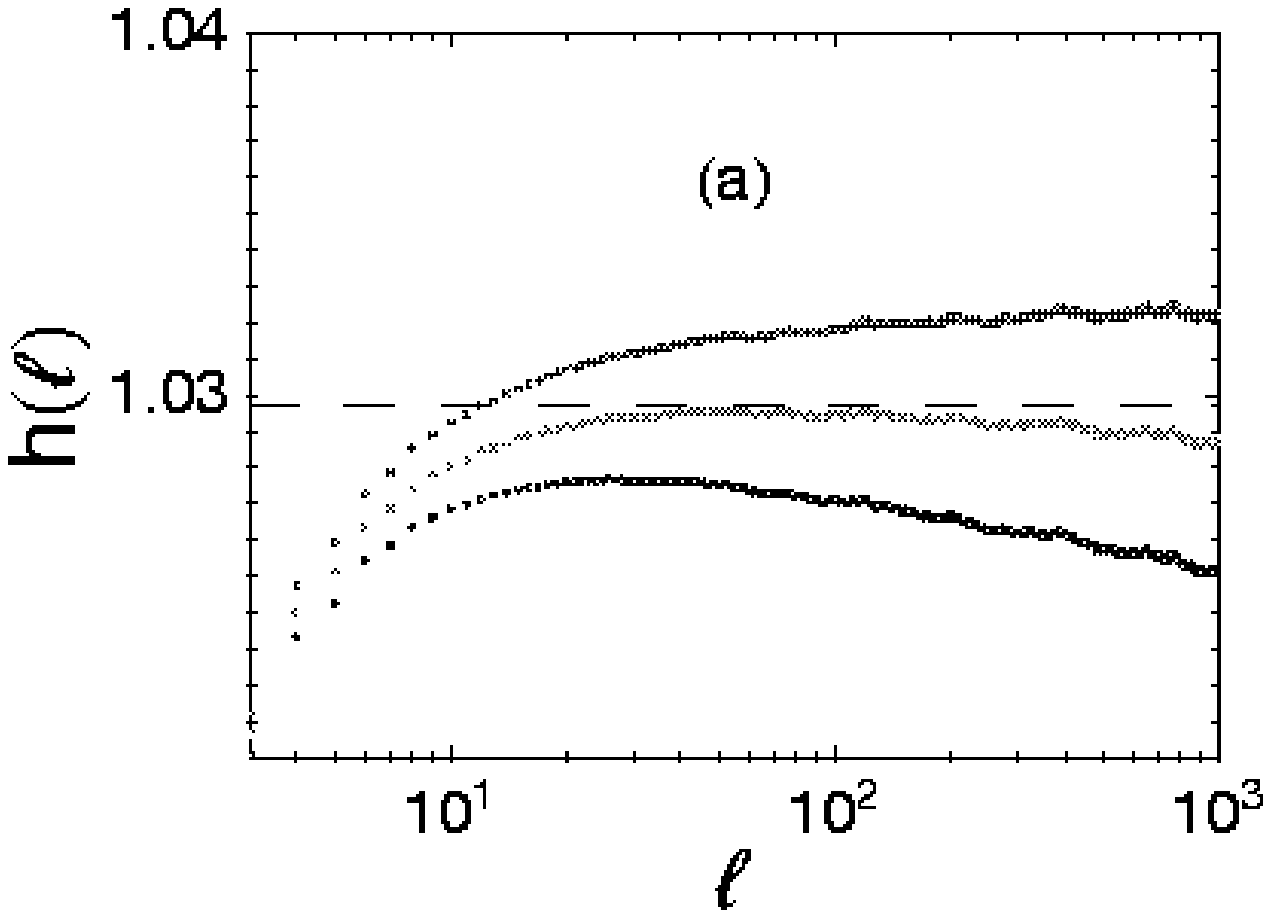}
}
\centerline{
\xsize
\epsfclipon
\epsfbox{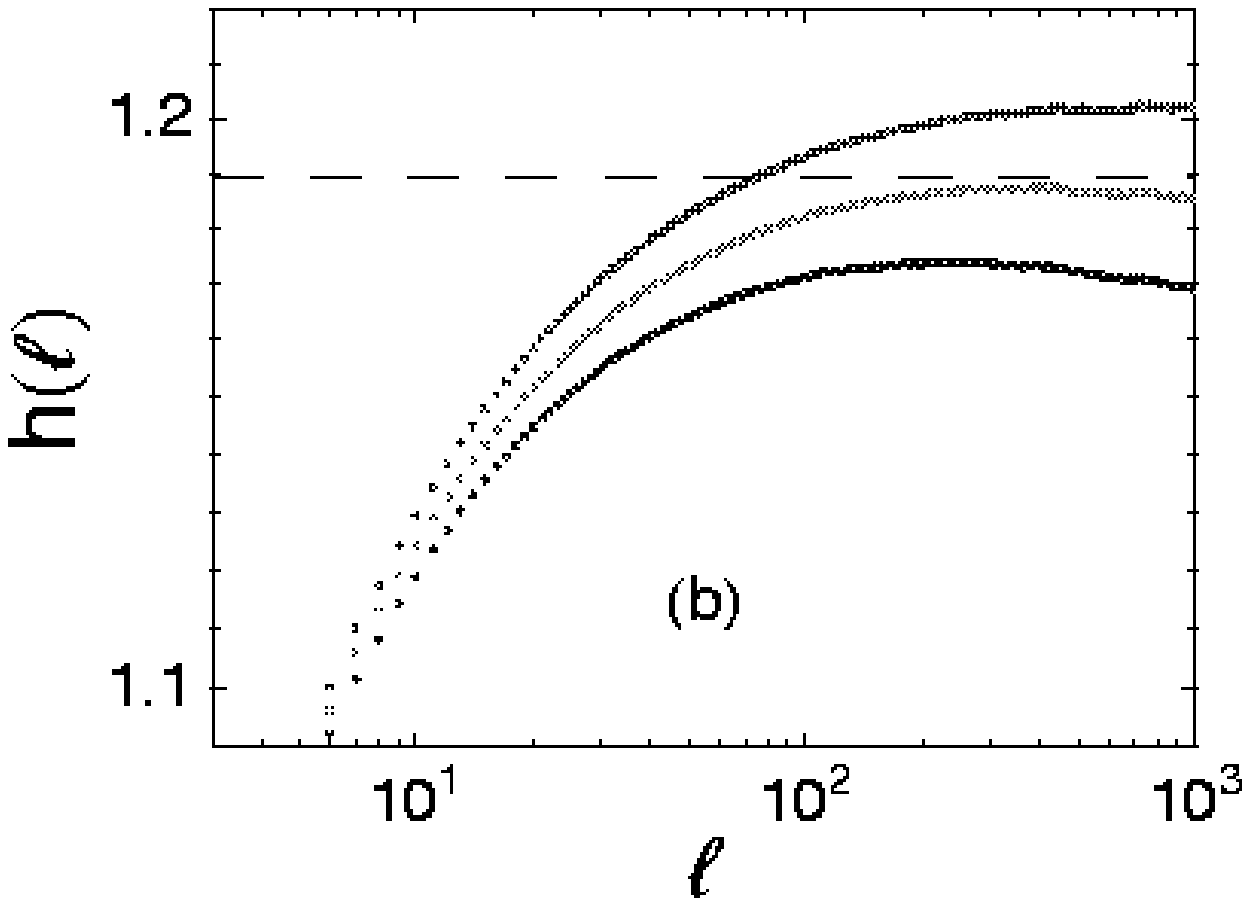}
}
\caption{$h(\ell)\equiv\langle r \rangle/\ell^z$ versus $\ell$ for (a) 5D
site percolation for values of (from top to bottom) $z'=0.5510$, 0.5515
and 0.5520, and (b) 5D bond percolation for values of (from top to
bottom) $z'=0.5595$, 0.5615, and 0.5635. The dashed horizontal lines are
provided as guides to the eye to allow one to better see that, for large
$\ell$, the the middle plots of $h(\ell)$ in (a) and (b) are
decreasing.}
\label{f5}
\end{figure}

Proceeding in the same manner for site and bond percolation in 5D (see
Fig.~\ref{f5}a,b), we find that the constant $A$ is positive for both
bond and site percolation, allowing us to determine only an upper bound
of $z=0.5515$ (the lower of the upper bounds for site and bond
percolation). While this method of finding bounds on $z$ by identifying
the value of $z'$ at which the slope of $h(\ell)$ changes sign does not
always yield both upper and lower bounds, it has the advantage that it
does not require any estimation of the parameters $A$ and $\Delta$ in
Eq.~(12) and, in fact, is somewhat insensitive to the exact form of the
the corrections-to-scaling terms.

\begin{figure}
\centerline{
\xsize
\epsfclipon
\epsfbox{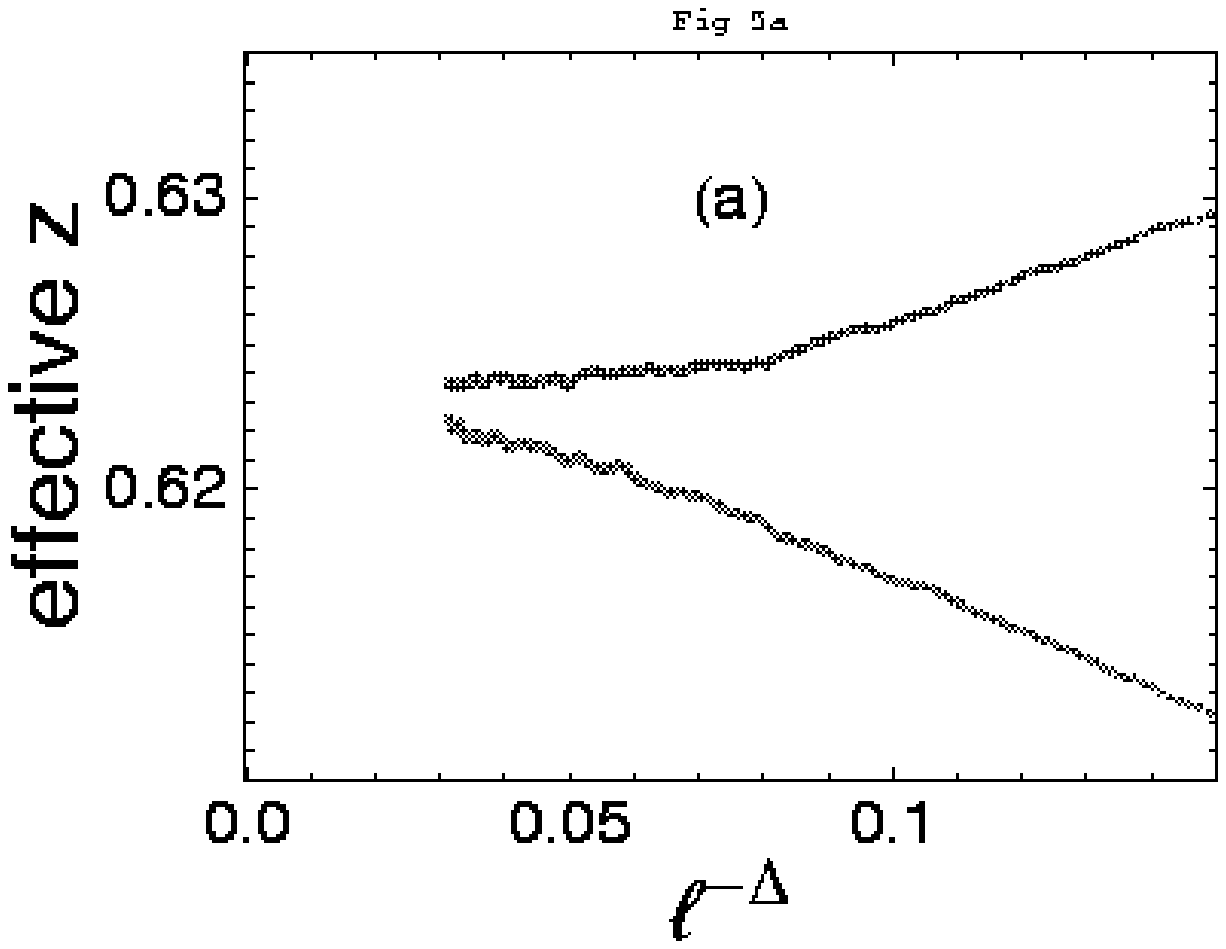}
} 
\centerline{
\xsize
\epsfclipon
\epsfbox{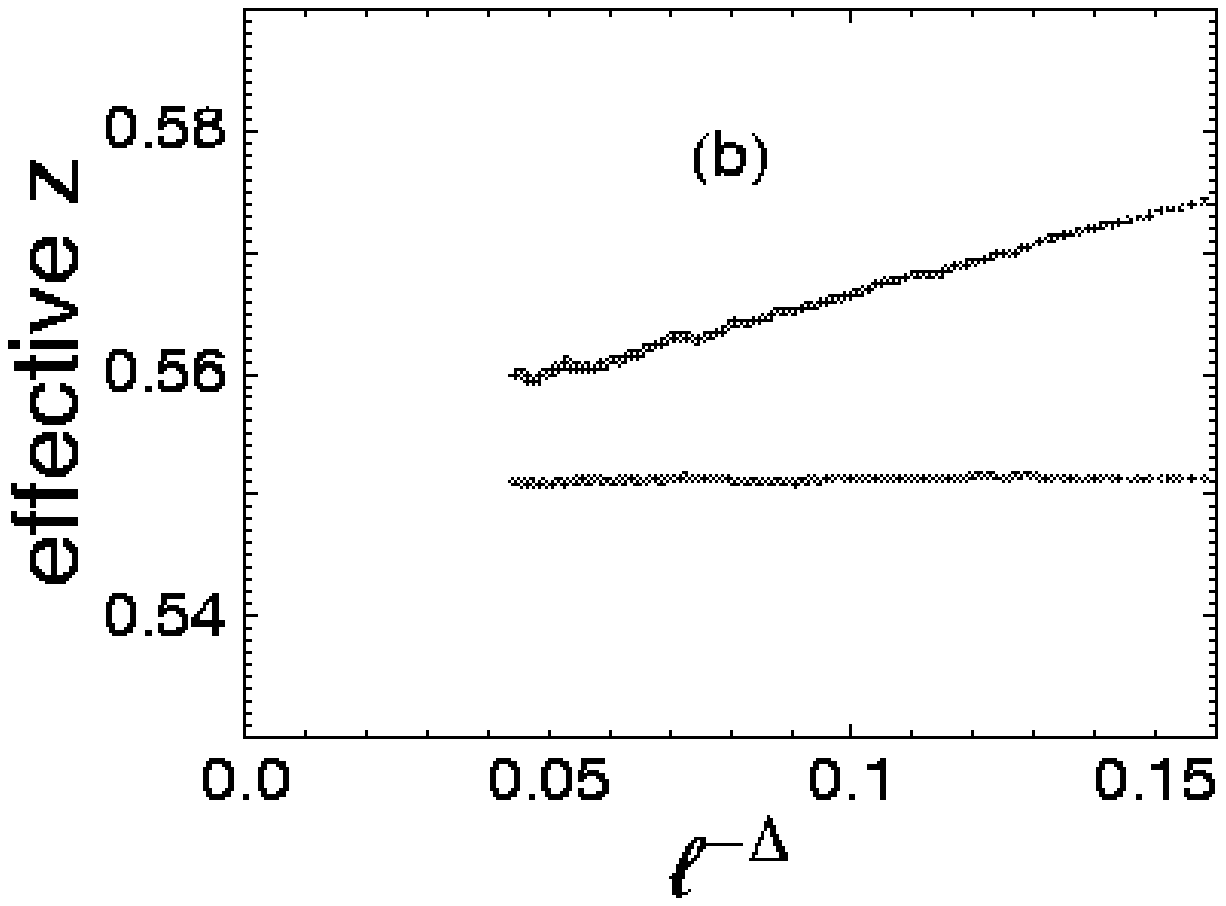}
}
\caption{Effective exponent $z$ versus $1/\ell^{-\Delta}$ with
$\Delta=0.5$ for bond percolation (upper line) and site percolation
(lower line) for (a) 4D and (b) 5D.}
\label{f6}
\end{figure}

We also analyze our data using another more commonly-used method
\cite{Grassberger92a,Grassberger92b,Ziff1984}. That method is to plot
the effective exponents, $z(\ell)$, between points $\ell$ and $2\ell$
versus $\ell^{-\Delta}$ using an estimated value of $\Delta$ which
yields the straightest line.  The effective exponent, $z(\ell)$, between
two point at $\ell$ and $2\ell$ is the value of the slope between these
points in a log-log plot of $\langle r(\ell)\rangle$
\begin{equation}
\label{e2x}
z(\ell)={\ln[\langle r(2\ell)\rangle]-\ln[\langle r(\ell)\rangle]\over
\ln[2\ell]-\ln[\ell]}={\ln[\langle r(2\ell)\rangle/\langle
r(\ell)\rangle]\over
\ln[2]}.
\end{equation}
The $\ell=0$ intercept of a plot of $z(\ell)$ will be an estimate for
$z$, and the slope will be proportional to $A$. Our best estimate for
$\Delta$ for $d=4$ and $d=5$ is $0.4<\Delta<0.6$, so we use a value of
$\Delta$ of 0.5 and plot $z(\ell)$ for 4D site and bond percolation in
Fig.~\ref{f6}a and 5D site and bond percolation in Fig.~\ref{f6}b. In
Fig.~\ref{f6}a, the fact that the slopes of the lines change suggests
that we are seeing the effects of both the correction-to-scaling term in
Eq.~(\ref{e12}) as well as higher order terms which become less
significant at larger values of $\ell$.  In general, it is more
efficient to generate smaller clusters and more of them, rather than
fewer, larger ones.  However, if the corrections-to-scaling are not well
understood or large, then one must build the largest clusters possible.
As we see here, strong corrections-to-scaling are present in 4D
percolation where the plots of effective slope change at large
$\ell$. If we had used smaller clusters using traditional
memory-management techniques we would have obtained incorrect results.
In Fig.~\ref{f6}a, the almost horizontal plot for site percolation
indicates that the amplitude, $A$, of the correction-to-scaling term is
very small.  From these plots and our estimates above of bounds on $z$,
we estimate 
\begin{equation}
\label{z4d5d}
z=\cases{
0.622\pm 0.002 & [4D]\cr
0.552\pm 0.002 & [5D]
}.
\end{equation}
In terms of $d_{\mbox{\scriptsize min}}$, this
corresponds to 
\begin{equation}
\label{dmin4d5d}
d_{\mbox{\scriptsize min}}=\cases{
1.607\pm 0.005 & [4D]\cr
1.812\pm 0.006 & [5D]
}.
\end{equation}
The previously published values for $d_{\mbox{\scriptsize min}}$ are 1.5
and 1.8 for 4D and 5D \cite{Stauffer}.  Thus our estimates of
$d_{\mbox{\scriptsize min}}$ are of considerably higher accuracy than
the existing ones and have accuracy comparable to that for the estimate
of $d_{\mbox{\scriptsize min}}$ in 3 dimensions, $1.374\pm 0.006$
\cite{Grassberger92a}.  Our results and all simulation parameters are
summarized in Table II.

\section{Discussion}

We have developed a technique which allows us to build very large
percolation clusters using very little memory. In fact, using the method
described here, relative to computer processing power available today
and in the foreseeable future, computer memory is no longer a constraint
on building percolation clusters near the percolation threshold. The
critical computer resource thus becomes solely processing power. For
example, by extrapolating from our simulations, we find that with our
method, with less than $10^8$ bytes of memory, we could build a 5D
cluster of $10^{12}$ sites, which would have required a lattice of
$10^{17}$ sites, and reach a value of $\ell_{\mbox{\scriptsize max}}$ of
$10^7$(versus the 1024 cutoff we used in our simulations).  But the time
to build a single trillion-site cluster would be about 2000 hours on
current workstations. As processor speeds increase, our technique for
reducing memory usage should allow critical exponents and constants to
be determined with greater precision.  Current techniques of growing
clusters, including the one described in this paper, require computer
processing resource of $O(s)$, where $s$ is the size of the cluster
grown.

We note that the technique we have developed is useful when we can count
the quantities in which we are interested as we build the cluster (e.g.,
cluster mass, average distance to sites in a chemical shell).  On the
other hand, it is not clear how we could calculate the mass of the
backbone, for example, using our method because current methods of
determining backbone mass require knowing all the sites in the cluster,
not just those in the current chemical shell.  To obtain backbone
properties one could, however, reduce memory required to $\sim s$
(versus $L^d$) by maintaining information about all visited sites (not
just those in the current chemical shell) in a data structure as opposed
to maintaining the full lattice data structure \cite{Vollmayer}.  In
Appendix A, we describe an alternative method of cluster generation
which can be used when information about all visited sites must be
maintained.

Finally, it is useful to compare our method with the Hoshen-Kopelman
method \cite{Hoshen}, which constructs {\it all\/} clusters in a
$d$-dimensional lattice by successively populating $d-1$ dimensional
slices of the lattice.  Memory is used to store the last and current
slice of the lattice so the memory needed scales as $L^{d-1}$.  The
Hoshen-Kopelman method is much less memory efficient than the method
presented here, and becomes less effective as the dimension increases
since $L^{d-1}/L^d\to 1$ with increasing $d$.  Also, the Hoshen-Kopelman
method cannot be used to calculate $d_{\mbox{\scriptsize min}}$.  On the
other hand, the Hoshen-Kopelman method is better suited to other
problems, such as calculating the number of clusters that span across a
rectangular system, than our method, based on the Leath algorithm.

\subsection*{Appendix A: Alternative Method of Cluster Growth}

We discuss a variant of our approach in which we still store information
concerning which sites are visited in a data structure and access the
entries using a hash table.  However, if one stores information about
all visited sites, not just for those in the last shell(s), then a
traditional random number generator (one which does not take the
coordinates/orientation of the bond as input) can be used. Let us first
consider the case where we have no need for occupied bond information
(e.g., we are simply counting the number of sites in the cluster). When
considering a growth site, if:

\begin{itemize} 

\item[{(i)}]
an adjacent site is vacant we determine whether the bond connected to
that site is occupied or not.

\item[{(ii)}]
an adjacent site is not vacant we simply do not make a determination of
whether the bond is occupied. 

\end{itemize} 

\noindent
In this way we make a determination about whether a given bond is
occupied no more than once. 

Now consider the case in which we do have a need to know whether a bond
is occupied or not (e.g., we are counting the number of bonds in the
cluster or we will be determining the backbone of the cluster). In this
case, when considering a growth site, if: 

\begin{itemize}
 
\item[{(i)}]
an adjacent site is vacant, we determine whether the bond connected to
that site is occupied or not. 

\item[{(ii)}]
an adjacent site is not vacant and is in a higher chemical shell, we
also determine whether the bond is connected to that site is occupied
or not. 

\item[{(iii)}]
an adjacent site is not vacant and is in the same chemical shell as the
growth site, we make a determination about whether the bond is occupied
only if the direction from the growth site to the adjacent site is
positive. In this way, the determination about whether the bond is
occupied is done only once. This situation arises in non-cubic
(e.g., triangular) lattices and when we start cluster growth with
multiple seeds. 

\item[{(iv)}]
an adjacent site is not vacant and is in a lower chemical shell than the
growth site, we make no determination about whether the bond to that
site is occupied; whether the bond is occupied has been determined
earlier in the growth process. In fact, the bond must be unoccupied
because if it were occupied we would have reached the growth site
earlier directly from the adjacent site. 

\end{itemize}

\noindent
Thus we ensure that we determine whether a bond is occupied once and
only once. If one needs to keep a record of whether a given bond is
occupied (e.g., to later determine the backbone) this information can be
stored in the entry in the data structure for the site with which the
bonds are associated along with the coordinates of the site, etc. 

This method trades off memory (we keep state for all visited sites)
versus performance (we can use the faster traditional random number
generators as opposed to the encryption-like random number generator). Also, in
cases where we, for some other reason, must keep state information about
all the sites, we can obtain the benefit of the using a faster random
number generator.

\subsubsection*{Acknowledgements}

We thank S. V. Buldyrev, D. Stauffer,  and Y. Ashkenazy for helpful discussions, and BP
Amoco and NSF for financial support.

\end{multicols}

\newpage

\begin{table}
\caption{Simulation parameters and results for $p_c$ and the Fisher exponent
$\tau$.\\ }
\begin{tabular}{|c|c|c|c|c|c|}
dimension & type & \# of realizations & $s_{\mbox{\scriptsize
max}}$ & $p_c$ & $\tau$ \\ \hline
  & bond & $10^8$ & 131~073 &$0.160130\pm 0.000003$ &  \\
4 & & & & & $2.313\pm 0.003$  \\
  & site & $10^8$ & 131~073 &$0.196889\pm 0.000003$ &  \\ \hline
  & bond & $10^8$ & 16~383 &$0.118174\pm 0.000004$ &  \\
5 & & & & & $2.412\pm 0.004$  \\
  & site & $10^8$ & 16~383 &$0.14081\pm 0.00001$ &  \\
\end{tabular}
\label{t1}
\end{table}

\begin{table}
\caption{Simulation parameters and results for the spreading exponent
$z$ and shortest path exponent $d_{\mbox{\scriptsize min}}$.\\ }
\begin{tabular}{|c|c|c|c|c|c|c|}
dimension & type & $p_c$ & \# of realizations & $\ell_{\mbox{\scriptsize
max}}$ & $z$ & $d_{\mbox{\scriptsize min}}$      \\ \hline
  & bond & 0.160130 & $73\times 10^6$ & 2048 & &  \\
4 & & & & & $0.622\pm 0.002$ & $1.607\pm 0.005$  \\
  & site & 0.196889 & $39\times 10^6$ & 2048 & &  \\ \hline
  & bond & 0.118174 & $105\times 10^6$ & 1024 & &  \\
5 & & & & & $0.552\pm 0.002$ & $1.812\pm 0.006$  \\
  & site & 0.14081 & $20\times 10^6$ & 1024 & &    \\
\end{tabular}
\label{t2}
\end{table}

\newpage

\newpage

%\begin{multicols}{2}

\np

\np

\np

%\end{multicols}


\begin{references}
\bibitem{text1}
The density of a cluster of size $L$ is $(1/L)^{d-d_f}$ where $d_f$ is
the fractal dimension. Since the co-dimension, $d-d_f$, increases with
$d$, the density decreases with $d$.
\bibitem{bondpercolatonused} 
We will use bond percolation in our discussion here. Application to site
percolation is straightforward.
\end{references}
\end{document}